# MEMS Mirror Manufacturing and Testing for Innovative Space Applications

**Alvise Bagolini** [1,4], **Maurizio Boscardin** [1,4], **Simone Dell'Agnello** [2], **Giovanni Delle Monache** [2], **Maurizio Di Paolo Emilio** [2], **Luca Porcelli** [2,3,*], **Lorenzo Salvatori** [2] and **Mattia Tibuzzi** [2]

[1] MicroNano Facility (MNF), Centre for Materials and Microsystems (CMM), Fondazione Bruno Kessler (FBK), Via Sommarive 18, Trento, Italy.

[2] Istituto Nazionale di Fisica Nucleare - Laboratori Nazionali di Frascati (INFN-LNF), Via E. Fermi 40, 00044, Frascati, Italy.

[3] Dipartimento di Fisica, Università della Calabria (Unical), Via Pietro Bucci, 87036, Arcavacata di Rende, Italy.

[4] Trento Institute for Fundamental Physics and Applications (TIFPA), Via Sommarive 14, 38126, Povo (TN), Italy.

* Corresponding Author: luca.porcelli@lnf.infn.it; Tel.: +39-06-9403-2733.

**Abstract:** In the framework of the GLARE-X (Geodesy via LAser Ranging from spacE X) project, led by INFN and funded for the years 2019-2021, aiming at significantly advance space geodesy, one shows the initial activities carried out in 2019 in order to manufacture and test adaptive mirrors. This specific article deals with manufacturing and surface quality measurements of the passive substrate of 'candidate' MEMS (Micro-Electro-Mechanical Systems) mirrors for MRRs (Modulated RetroReflectors); further publications will show the active components. The project GLARE-X was approved by INFN for the years 2019-2021: it involves several institutions, including, amongst the other, INFN-LNF and FBK. GLARE-X is an innovative R&D activity, whose at large space geodesy goals will concern the following topics: inverse laser ranging (from a laser terminal in space down to a target on a planet), laser ranging for debris removal and iterative orbit correction, development of high-end ToF (Time of Flight) electronics, manufacturing and testing of MRRs for space, and provision of microreflectors for future NEO (Near Earth Orbit) cubesats. This specific article summarizes the manufacturing and surface quality measurements activities performed on the passive substrate of 'candidate' MEMS mirrors, which will be in turn arranged into MRRs. The final active components, to be realized by 2021, will inherit the manufacturing characteristics chosen thanks to the presented (and further) testing campaigns, and will find suitable space application to NEO, Moon, and Mars devices, like, for example, cooperative and active lidar scatterers for laser altimetry and lasercomm support.

## 1. Introduction

The SCF_Lab (Satellite/lunar/GNSS laser ranging/altimetry and cube/microsat Characterization Facilities Laboratory) is an infrastructure devoted to space R&D, which has been operational for the last 15 years. It belongs to INFN-LNF, and it has got widely recognized capabilities for the design, test and space qualification of space laser retroreflectors, and laser ranging systems and services. The laboratory itself is located within an 85 $m^2$ class 10,000 (ISO 7) clean room, with separate entry areas for operators and equipment. The clean room hosts two OGSEs (Optical Ground Support Equipments): respectively, the SCF, and the SCF-G (especially thought for satellite navigation applications) apparata. Each one of the OGSE can independently run thermo-optical qualification tests, thanks to a substantial dotation of hardware, including: cryostat, positioning system, thermal control system, control electronics, vacuum system, AM0 solar simulator, optical bench with imagers, and laser wavefront Fizeau interferometer (Figure 1).

The SCF_Lab has developed and refined, over the years, the crucial ability of concurrent measure and modelling of CCR's optical FFDP (Far Field Diffraction Pattern), and temperature distribution of laser ranging LRAs (Laser Retroreflector Arrays) in a laboratory-simulated space environment w.r.t. temperature, vacuum and solar constant (through the AM0 solar simulator) [1]. Over the last 15 years, several topical qualification activities were performed, including, yet not limited to, what is described in [1-7].

It is worth pointing out that part of the 'candidate' MEMS mirrors under testing may contribute to the realization (or be itself part) of actual flight hardware for future missions; any eventual space qualification activity will be described in further works, and it is already well within reach of the capabilities of the SCF_Lab, as per [5-7].

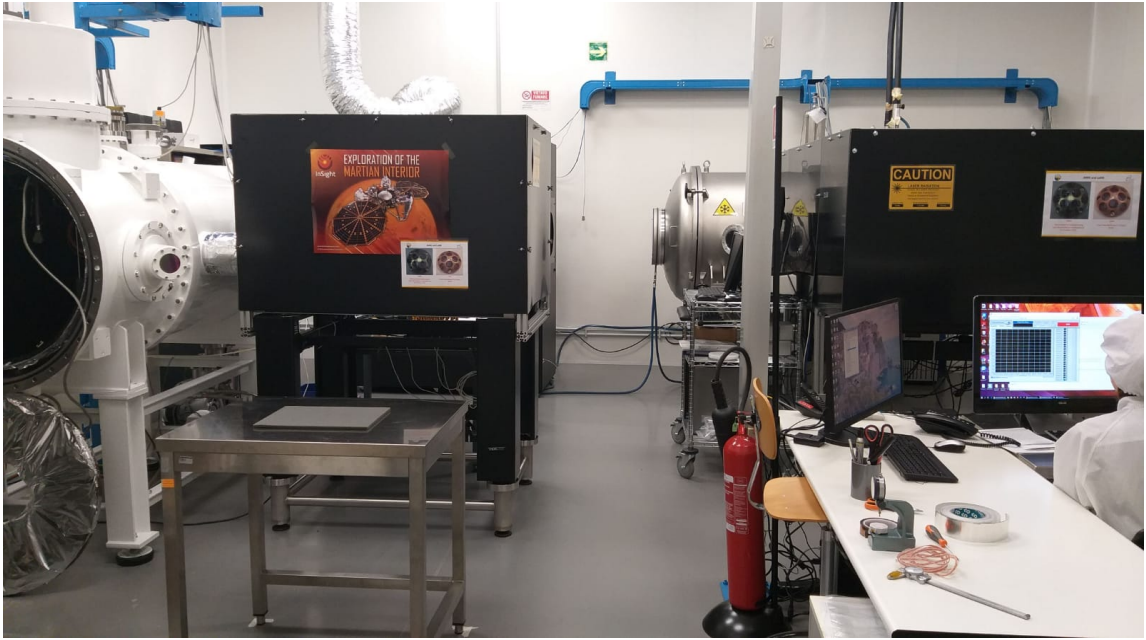

**Figure 1.** The SCF_Lab. From left to right: the SCF climatic chamber (white cryostat, whose open porthole faces the first AM0 solar simulator), the SCF optical bench (black box), the SCF-G climatic chamber (steel cryostat, whose closed porthole faces the second AM0 solar simulator), the SCF-G optical bench (black box), and the operator control stations.

FBK, the other participant in this activity, is a research non-profit public interest entity. It comprises seven research centers with 410 researchers, two specialized libraries and seven laboratories. MNF is the microfabrication and characterization facility of FBK. It provides the entire development cycle of advanced miniaturized devices for both R&D and production, based on more than 20 years of experience, in a 650 m$^2$ class ISO4-5, ISO9001 certified clean room facility. MNF's main fields of expertise are: Silicon Radiation Sensors, MEMS, BioMEMS, Advanced Photonics & Photovoltaics.

The following Section 2 will describe the passive substrate manufacturing of the 'candidate' MEMS mirrors under testing (carried out by MNF), and the subsequent surface quality measurements activities of the same mirrors (carried out by the SCF_Lab).

## 2. Materials and Methods

*2.1. Mirror Manufacturing*

Mirrors were fabricated on 6 inches SEMI[1] standard single crystal silicon wafers, using standard semiconductor IC (Integrated Circuit) processing technology. They were organized in two groups, one (group A) to test photoresist as sacrificial layer, the other (group B) to test a silicon oxide sacrificial layer. The sacrificial layer is located under the mirror layer: in the complete device fabrication, it is meant to be selectively removed from under the mirror, to release it and allow electro-mechanical actuation. Being under the mirror layer, the sacrificial layer may alter its performance, affecting its roughness and waviness.

Group A mirrors (3 overall) were fabricated by spin coating standard photoresist with thickness of 1200 nm, and then evaporating 500 nm aluminum by e-beam CVD (Chemical Vapor Deposition).

Group B mirrors (5 overall) consisted of a silicon oxide with thickness of 1000 nm, covered with 500 nm RF (Radio Frequency) sputtered aluminum. The silicon oxide was either grown by thermal diffusion (samples 1 to 4) or TEOS (tetraethyl orthosilicate) deposited by LPCVD (Low Pressure CVD, sample 5). All aluminum layers were deposited at room temperature, except for samples 3 and 4, for which the sample temperature was raised to 150 °C during aluminum RF sputtering. Higher aluminum deposition temperature makes the layer more robust towards further fabrication steps such as photolithography, which induces a thermal load on the wafer, but raising the deposition temperature may also increase the mirror roughness.

The following Table 1 summarizes the mirror sample characteristics.

**Table 1.** Mirror sample characteristics.

| Group[1] | Fabrication Steps[2] | Layer Thickness [nm] | Wafer Checklist[3] 1 | 2 | 3 |
|---|---|---|---|---|---|
| A (3 mirrors) | 1 (photoresist standard) |  | x | x | x |
|  | 2 (Al e-beam CVD) | 500 | x | x | x |
|  | 3 (process load simulation[4]) |  | x |  |  |

| Group[1] | Fabrication Steps[2] | Layer Thickness [nm] | Wafer Checklist[3] 1 | 2 | 3 | 4 | 5 |
|---|---|---|---|---|---|---|---|
| B (5 mirrors) | 1 (thermal oxide) | 1000 | x | x | x | x |  |
|  | 2 (TEOS) | 1000 |  |  |  |  | x |
|  | 3 (Al RF sputtering @ RT[5]) | 500 | x | x |  |  | x |
|  | 4 (Al RF sputtering @ 150 °C) | 500 |  |  | x | x |  |
|  | 5 (process load simulation[4]) |  | x |  | x |  |  |

[1] Group A totals 3 mirrors, and Group B totals 5 mirrors. Overall, 8 samples were investigated. [2] Group A needs 3 fabrication steps, while Group B needs 5 fabrication steps. [3] Given the different fabrication steps of each group, each wafer underwent diverse production flow, which is correspondingly checked in the table. [4] It is the simulation of the manufacturing process thermal load. [5] RT (Room Temperature) ~ 20 °C.

Concerning the selection of the appropriate metal as mirror layer, the potential and envisaged space deployment of the mirrors was considered, taking into account:
- Existing literature, especially what was generated by space agencies [8-9].
- Everyday best practices and lessons learnt at the SCF_Lab [1-7].

Hence, the final preference for the mirror layer was given to aluminum because of the following:

---

[1] www.semi.org.

- W.r.t. silver: aluminum does not need an extra protective coating to avoid oxidation.
- W.r.t. gold: due to its higher solar absorptivity, gold is not recommend when dealing with space applications.

Figure 2 shows an example of mirror, which was manufactured by FBK and whose surface quality was tested at the SCF_Lab.

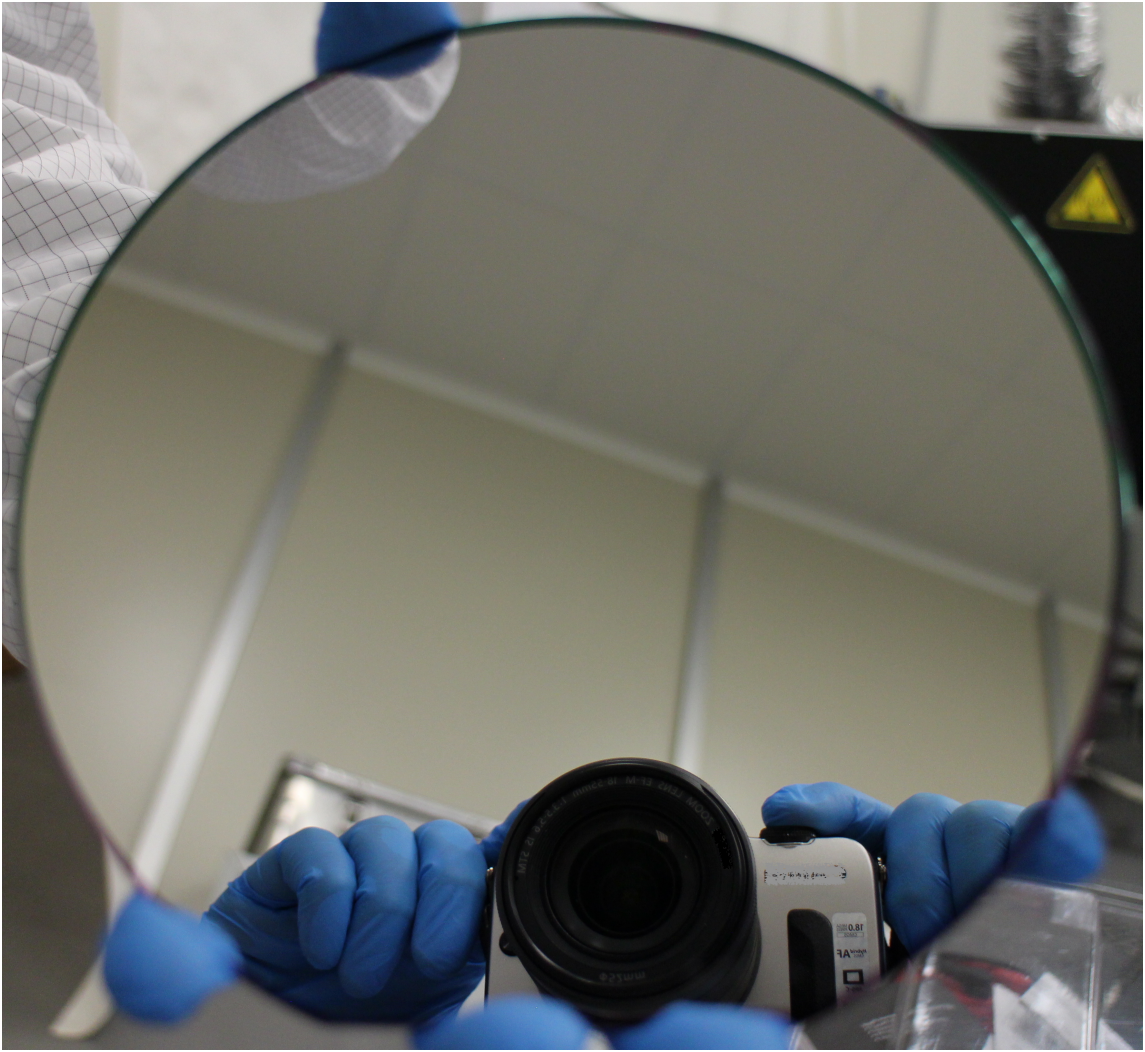

**Figure 2.** The picture shows an example mirror, which was manufactured by FBK and whose surface quality was tested at the SCF_Lab. The diameter of the mirrors is 6 inches.

*2.2. Optical Quality Measurements of the Mirrors*

During this first optical features assessment, carried out at the SCF_Lab, of the passive substrate of the 'candidate' MEMS mirrors, measurements were performed in the near field regime, through a 4D Technology AccuFiz wavefront Fizeau interferometer. The far field campaign, which is the 'house specialty' of the SCF_Lab (as per [1-7]), will be shown in a following contribution.

For this kind of in-house designed, developed, and manufactured optics, it was necessary to 3D-print a custom interface for integration with standard optical mounts (Figure 3). The need for a 3D-printed plastic holder is also dictated by the fact that these 'candidate' MEMS mirrors will be actuated: we would like to keep them in electrical discontinuity w.r.t. the rest of the metallic parts of the optical bench.

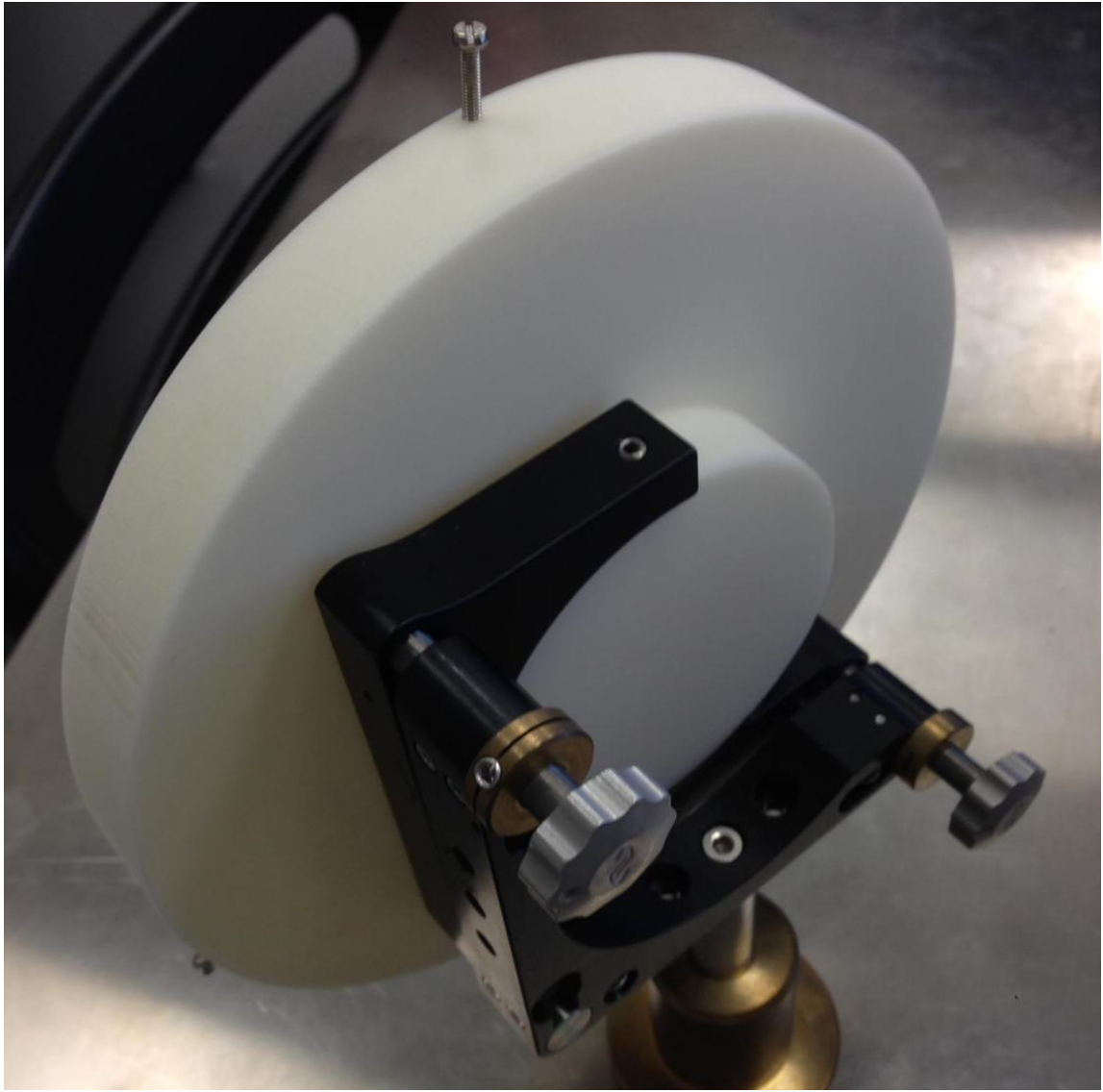

**Figure 3.** 3D-printed custom interface (white disks) for integration with standard optical mounts (aluminum, black, and brass parts).

The near field campaign consisted in the reflected wavefront characteristics measurements, performed through a Fizeau interferometer, which is part of the hardware dotation of the SCF_Lab. As per following Section 3, thanks to the interferometer, it is possible to measure the near field wavefront reflected by any mirror under investigation, and which brings with it the optical 'fingerprints' of the tested mirror itself. The comparison between the ideal plane wavefront emitted by the interferometer and the 'slightly' aberrated one retroreflected by the mirror allows the mining of quantitative information about the quality of optical surfaces.

## 3. Near Field Campaign Results

Tables 2 and 3 summarize the measured mirror sample optical features. Figure 4 shows an excerpt of the mirror positioning on the optical bench. For each 'candidate' MEMS we analysed 1 interferogram, which was, in turn, produced averaging out over 40 shots, taken with a centered 130-mm wide acquisition mask, in order to cut out edge effects originating from the larger diameter (Figure 2). These 8 average interferograms (one per mirror) were analysed with 3 centered analysis masks of different diameter: 33 mm, 65 mm, and 130 mm. We extracted the following information for each mirror/diameter at the operational wavelength of the interferometer (633 nm):

- The PV (Peak-to-Valley) error, which is measured comparing the reflected wavefront with the non-aberrated one emitted by the interferometer itself (Tables 2 and 3, second to last columns). This quantity brings information about the 'worst' possible waviness of the surface of interest, averaged out over the whole analysed area. The associated measurement error is ± 0.1 wv.
- The RMS (Root Mean Square) error, which is similar to the PV one, but it provides information about how smooth a wavefront is on average and 'locally', in a statistical sense (Tables 2 and 3, last columns). The associated measurement error is ± 0.01 wv.
- A 'qualitative' FFDP, computed through the interferometer software in arbitrary units, which will have to be compared with the one measured on the SCF_Lab optical bench, at 532 nm, in a following publication. As an example, only the 'qualitative' FFDPs of Group A's mirror number 3 are shown, in the following thumbnails (Figure 5).

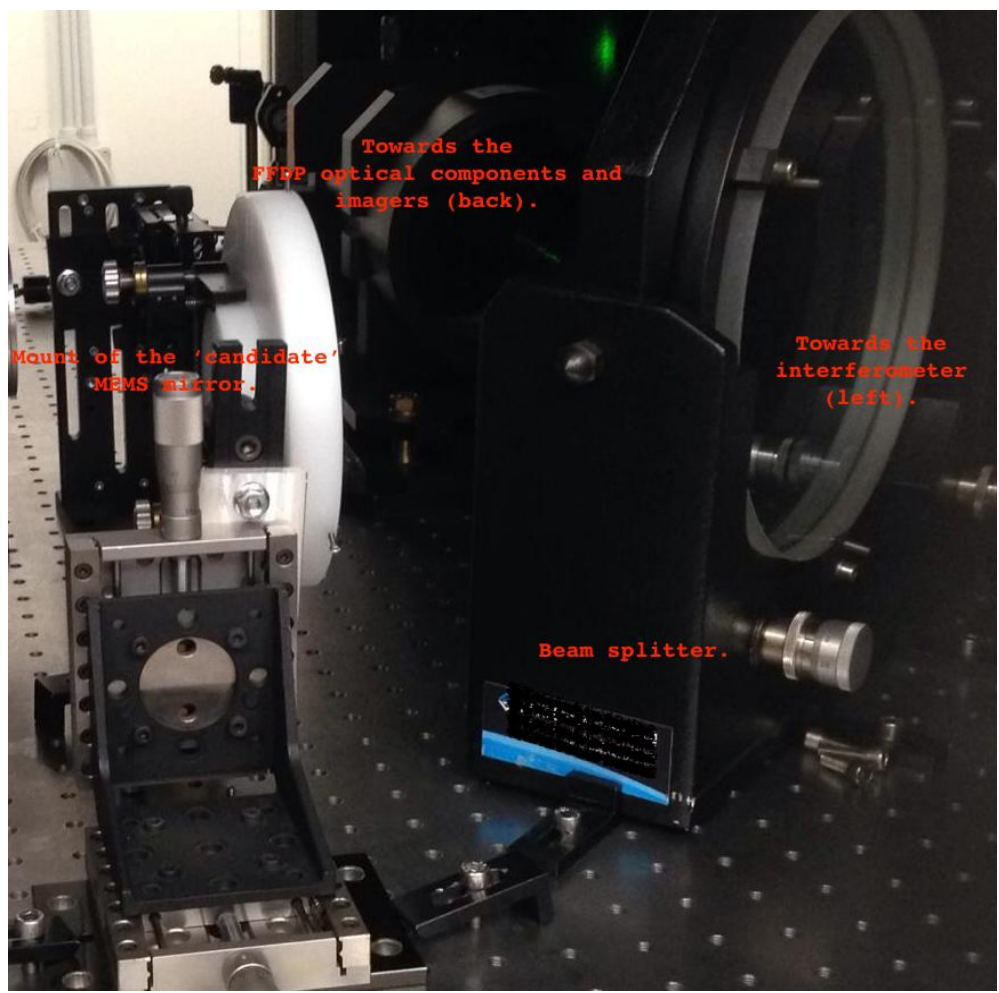

**Figure 4.** Excerpt of the mirror positioning on the optical bench.

Table 2. Measured Group A mirror sample optical features.

| Group[1] | Wafer Number[2] | Mask Diameter[3] [mm] | PV [wv] | RMS [wv] |
|---|---|---|---|---|
| A (3 mirrors) | 1 | 33 | 0.4 | 0.06 |
| | | 65 | 1.7 | 0.29 |
| | | 130 | 9.5 | 1.51 |
| | 2 | 33 | 0.5 | 0.10 |
| | | 65 | 3.0 | 0.49 |
| | | 130 | 28.7 | 3.26 |
| | 3 | 33 | 0.2 | 0.02 |
| | | 65 | 1.3 | 0.17 |
| | | 130 | 11.3 | 1.33 |

[1] Group A totals 3 mirrors. [2] Group A wafer numbers are 1, 2, and 3. [3] The diameter of the analysis mask used to extract information from the raw interferograms. For the 130-mm diameter case only, acquisition mask and analysis mask are the same, in order to cut out edge effects.

Table 3. Measured Group B mirror sample optical features.

| Group[1] | Wafer Number[2] | Mask Diameter[3] [mm] | PV [wv] | RMS [wv] |
|---|---|---|---|---|
| B (5 mirrors) | 1 | 33 | 0.9 | 0.17 |
| | | 65 | 3.1 | 0.42 |
| | | 130 | 13.7 | 2.24 |
| | 2 | 33 | 0.9 | 0.17 |
| | | 65 | 3.1 | 0.50 |
| | | 130 | 13.8 | 2.20 |
| | 3 | 33 | 1.4 | 0.18 |
| | | 65 | 6.5 | 1.14 |
| | | 130 | 67.3 | 7.27 |
| | 4 | 33 | 1.0 | 0.15 |
| | | 65 | 44.6 | 5.36 |
| | | 130 | 88.60 | 15.80 |
| | 5 | 33 | 1.0 | 0.15 |
| | | 65 | 4.4 | 0.73 |
| | | 130 | 28.0 | 4.83 |

[1] Group B totals 5 mirrors. [2] Group B wafer numbers are 1, 2, 3, 4, and 5. [3] The diameter of the analysis mask used to extract information from the raw interferograms. For the 130-mm diameter case only, acquisition mask and analysis mask are the same, in order to cut out edge effects.

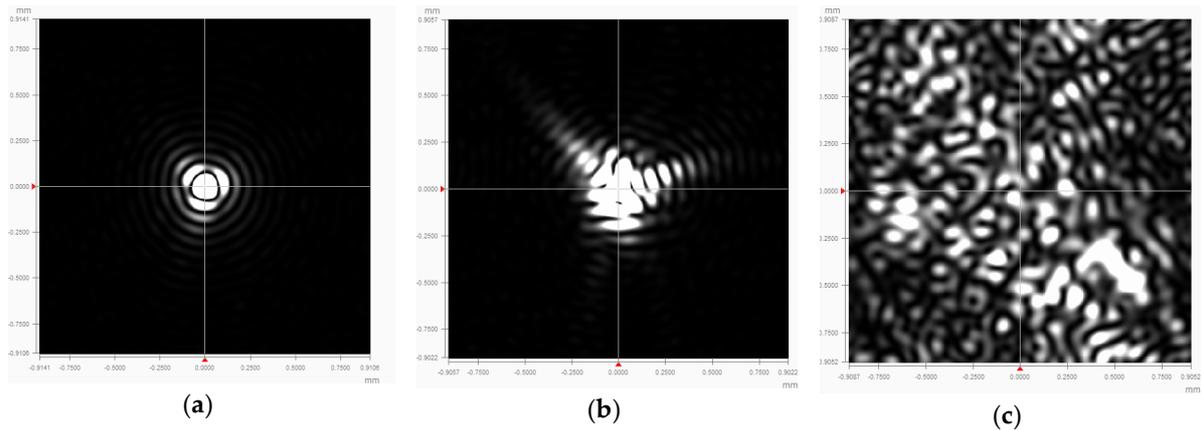

|  (a) | (b) | (c) |

**Figure 5.** The 3 'qualitative' FFDPs of Group A's mirror number 3, computed through the interferometer software in arbitrary units, at the operational wavelength of 633 nm: (**a**) when the FFDP is computed with a 33-mm wide analysis mask, the PV and RMS errors of the corresponding subsampled interferogram are small (as per Table 2), and the diffraction pattern closely resembles an Airy distribution; (**b**) the bigger the analysis mask (65 mm in this case), the bigger the PV and RMS errors (as per Table 2), and the farther away from an Airy distribution the diffraction pattern (PV ≈ 3 wv is the global waviness error for which the diffraction pattern starts being severely distorted); (**c**) when the PV error is of the order of 10 wv or more, the diffraction pattern is compromised, and the reflected photons are scattered all over the detector area and beyond.

## 4. Discussion and Conclusions

In the framework of the GLARE-X project, this article shows the initial activities carried out in order to manufacture and test adaptive mirrors. In particular, one has shown the manufacturing and surface quality measurements of the passive substrate of 'candidate' MEMS mirrors for MRRs. During this first optical features assessment, the measurements were performed in the near field regime, through a 4D Technology AccuFiz wavefront Fizeau interferometer.

Putting together the manufacturing information provided in Section 2 with the measured data of Tables 2 and 3, and Figure 5 (and alike), one can reach several first assessment conclusions:
- There seems to be a distinctive fabrication-dependent waviness: Group A's PVs and RMS's are consistently smaller than Group B's corresponding quantities.
- There seems to be a clear dimension-dependent waviness: PVs and RMS's errors of all the subsampled interferogram (with analysis mask diameters of 33 mm and 65 mm) are smaller, and all the diffraction patterns closely resemble Airy distributions (we showed only Group A's mirror number 3 FFDPs in Figure 5).
- As foreseen in Section 2, Group B's samples 3 and 4, for which the sample temperature was raised to 150 °C during manufacturing, show the biggest values overall for both PV and RMS.

The outcome of the present first optical features assessment seems to suggest that photoresist has to be preferred to silicon oxide as a sacrificial layer for the 'actual' MEMS mirrors for MRRs (to be realized by 2021), in order to minimize the waviness of the optical surfaces. As a thumb rule, the lower the waviness (both PV and RMS), the more suitable the surface to be used as an active laser ranging terminal in space for the kind of high-end envisaged activities [10]. However, the far field campaign, and following thermo-optical qualification tests will determine the final choices for the space qualified hardware.

Today, laser communications systems are particularly used for free space satellite data communication. An optical system based on MRR architecture could have a strong impact on time-to-market of satellites by increasing the communication data rate in view of future envisaged space applications, such as high resolution cartography/geodesy of the Moon and Mars, where a high and efficient ability to transmit large volumes of data is required (i.e., 10-cm resolution planetary maps). In fact, recently, space agencies, research institutes, and companies have been working hard to increase their efforts to design and manufacture reliable MRR devices for optical space communications [11-15].

As already described, in the following testing activities, we will characterize active optical components (MEMS), which will be subsequently assembled into retroreflectors (MRRs). The GLARE-X project will therefore demonstrate the ability of its participants to create a complete MRR production chain, including design, manufacturing, thermo-vacuum-optical testing, and qualification for space flight.

**Funding and Acknowledgments:** The authors would like to thank INFN and the Autonomous Province of Trento for supporting the present R&D effort on active laser retroreflectors payloads through the INFN's National Scientific Committee n. 5 and the INFN-FBK Convention.

**References**

1. Dell'Agnello, S., Delle Monache, G., Currie, D. G., et al. Creation of the new industry-standard space test of laser retroreflectors for the GNSS and LAGEOS. *Advances in Space Research* **2011**, 47, 822-842, doi:10.1016/j.asr.2010.10.022.
2. Dell'Agnello, S., Delle Monache, G., Currie, D. G., et al. ETRUSCO-2: an ASI-INFN project of technological development and SCF-Test of GNSS laser retroreflector arrays, in *ESA Proceedings of the 3rd International Colloquium - Scientific and Fundamental Aspects of the Galileo Programme*, Copenhagen, Denmark, 2011.
3. Dell'Agnello, S., Boni, A., Cantone, C., et al. Thermo-optical vacuum testing of Galileo In-Orbit Validation laser retroreflectors. *Advances in Space Research* **2016**, 57, 2347-2358, doi:10.1016/j.asr.2016.03.025.
4. Porcelli, L., Boni, A., Ciocci, E., et al. Thermo-optical vacuum testing of IRNSS laser retroreflector array qualification model. *Advances in Space Research* **2017**, 60, 1054-1061, doi:10.1016/j.asr.2017.05.012.
5. Dell'Agnello, S., Delle Monache, G., Porcelli, L., et al., INRRI-EDM/2016: the first laser retroreflector on the surface of Mars. *Advances in Space Research* **2017**, 59, 645-655, doi:10.1016/j.asr.2016.10.011.
6. Dell'Agnello, S., Delle Monache, G., Ciocci, E., et al. LaRRI: Laser Retro-Reflector for InSight Mars Lander. *Space Research Today* **2017**, 200, 25-32.
7. Porcelli, L., Tibuzzi, M., Mondaini, C. et al. Optical-Performance Testing of the Laser RetroReflector for InSight. *Space Science Reviews* **2019**, 215:1, doi:10.1007/s11214-018-0569-3.
8. Henninger, J. H. *Solar Absorbance and Thermal Emittance of Some Common Spacecraft Thermal-Control Coatings*, NASA Reference Publication 1121, 1984.
9. Echániz Ariceta, T. Infrared spectral emissivity studies on metals and materials for solar thermal applications. PhD Thesis, Basque Country University, Leioa, Spain, March 2016.
10. Dell'Agnello, S., Delle Monache, G., Porcelli, L., et al. Testing Gravity with the Moon and Mars, in *Proceedings of the Conference Frontier Objects in Astrophysics and Particle Physics*, Frascati Physics Series Vol. 66 (2018), Vulcano, Italy, 2018.
11. Ziph-Schatzberg, L., Bifano, T., Cornelissen, S., et al. Secure optical communication system utilizing deformable MEMS mirrors, in *Proceedings of SPIE, Volume 7209, MEMS Adaptive Optics III, 72090C (23 February 2009)*, doi: 10.1117/12.812145.
12. Boston Micromachines Corporation. Optical Modulator Technical Whitepaper. www.bostonmicromachines.com, February 2011.
13. Chopard, A. Demonstration of free space optical communication link in 10-meters range using electro-absorption modulator arrays. Master Thesis, Kungliga Tekniska Högskolan, Stockholm, Sweden, March 2017.
14. Guillen Salas, A., Stupl, J., and Mason, J. Modulating Retro-Reflectors: Technology, Link Budgets and Applications, in *Proceedings of the 63rd International Astronautical Congress*, Naples, Italy, 2012.
15. ESA ITT AO/1-7757/14/NL/SW, 2014. Planetary communication system based on modulated retro-reflection.